\documentclass[twocolumn]{aa}
\usepackage{graphicx}
\usepackage{psfig}
\usepackage{graphicx}
\begin{document}

\title{Evidence for a co-moving sub-stellar companion of GQ Lup \thanks{Based on
observations obtained on Cerro Paranal, Chile, in ESO programs 73.C-0164 
and 273.C-5047 as well as on data collected at the Subaru Telescope 
and the Hubble Space Telescope, both obtained from their science archives.} }

\author{R. Neuh\"auser \inst{1}, E.W. Guenther \inst{2}, G. Wuchterl \inst{1},
M. Mugrauer \inst{1}, A. Bedalov \inst{1} \and P.H. Hauschildt \inst{3}}

\offprints{R. Neuh\"auser, rne@astro.uni-jena.de}

\institute{Astrophysikalisches Institut, Universit\"at Jena,
Schillerg\"a{\ss}chen 2-3, D-07745 Jena, Germany 
\and Th\"uringer Landessternwarte Tautenburg, Sternwarte 5, D-07778 Tautenburg, Germany 
\and Hamburger Sternwarte, Gojenbergsweg 112, D-21029 Hamburg, Germany }

\date{received Oct 2004; accepted March 2005}

\abstract{We present a companion of the $\le 2$ Myr 
young classical T Tauri star 
GQ Lup in the Lupus star forming region at $140 \pm 50$ pc
from imaging, astrometry, and spectroscopy. With direct K-band imaging using VLT/NACO, 
we detected an object 6 mag fainter than GQ Lup located $0.7^{\prime \prime}$ west of it.
Compared to images obtained 2 to 5 years earlier with Subaru/CIAO and HST/PC, 
this object shares the proper motion of GQ Lup by 5 and $7\sigma$, respectively, 
hence it is a co-moving companion.
Its $K-L'$ color is consistent with a spectral type early to mid L. 
Our NACO K-band spectrum yields spectral type 
M9-L4 with H$_{2}$O and CO absorption,
consistent with the new GAIA-Dusty template spectrum
for $\log~g \simeq 2$ to 3
and T$_{\rm eff} \simeq 2000$ K 
with $\sim 2$~R$_{\rm jup}$ radius at $\sim 140$ pc, hence few Jupiter masses.
Using the theoretical models from Wuchterl \& Tscharnuter (2003),
Burrows et al. (1997), 
and Baraffe et al. (2002),
the mass lies between 1 and 42 Jupiter masses.
}

\maketitle

\markboth{Neuh\"auser et al.: A sub-stellar companion of GQ Lup}{}

\section{Introduction: Direct detection of planets}

Direct detection of sub-stellar companions is difficult because of the 
large dynamic range between the faint companion and the close-by, 
much brighter star. Due to contraction, Myr young sub-stellar objects are 
brighter than Gyr older ones. 
Hence, young nearby stars would be the best targets 
for the direct detection of sub-stellar companions.
The mass of a detected sub-stellar companion can be estimated 
from the observed companion magnitude and the assumed 
or known age and distance of the primary star, using theoretical model calculations.
Chauvin et al. (2004) reported the detection of a faint object near the 
brown dwarf 2M1207, a potential member of TWA at $\sim 8$ Myrs, 
whose infrared (IR) color and H-band spectrum are consistent with an L dwarf,
hence $\sim 5$~M$_{\rm jup}$ (Baraffe et al. 2002 models), 
if bound (astrometry missing).

Here, we present evidence for a sub-stellar companion around GQ Lup
located in the Lupus I cloud (Tachihara et al. 1996), a 
young K7eV-type classical T Tauri star with low extinction, 
but with mid- and far-IR excess (Hughes et al. 1994), 
i.e. a disk, and also both soft and hard X-ray emission
(Krautter et al. 1997), rare for classical 
T Tauri stars (Neuh\"auser et al. 1995), 
so that we included this star in our planet search programs
by radial velocity and direct imaging since March 1999.
As distance towards GQ Lup in the Lupus I cloud,
we use $140 \pm 50$ pc (Wichmann et al. 1998, 
Neuh\"auser \& Brandner 1998, Knude \& Hog 1998),
the age of GQ Lup is $\sim 0.1$ to 2 Myrs,
depending on the set of models used
(we work with $1 \pm 1$ Myr).

\section{Direct observations of a wide companion}

\subsection{AO imaging detection and photometry}

We observed GQ Lup with Yepun (ESO-VLT-UT4) on Cerro Paranal, Chile,
using the Adaptive Optics (AO) instrument NACO (Naos-Conica, Rousset et al. 2003)
on 25 June 2004 in visitor mode using the S13 camera 
($14^{\prime \prime} \times 14^{\prime \prime}$ field) in the K$_{\rm s}$-band.
We took 200 co-adds $\times 0.347s \times 27$ jitter positions.
Data reduction was done with {\it eclipse}: dark subtraction,
flat devision, shift+add.
A companion candidate is found $0.7^{\prime \prime}$ west of GQ Lup (Fig. 1).
By comparison with the primary (K=$7.096 \pm 0.020$ mag, 2MASS),
we got K$_{\rm s} = 13.1 \pm 0.1$ mag for the companion candidate
with aperture photometry after subtracting the primary.

\begin{figure}
\vbox{\psfig{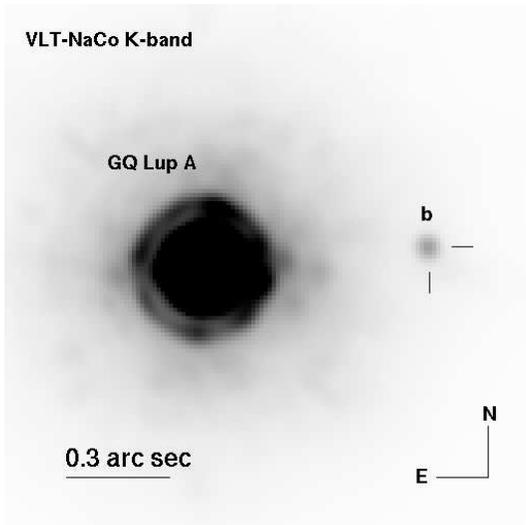}}
\caption{VLT-NACO K$_{\rm s}$-band image of GQ Lup and its
6 mag fainter companion candidate $0.7325 \pm 0.0034^{\prime \prime}$ west.}
\end{figure}

K- and L'-band images of GQ Lup were obtained with the 
Subaru Coronagraphic Imager with Adaptive Optics (CIAO, 
Murakawa et al. 2004),
in program o02312, retrieved by us from the public Subaru archive SMOKA.
We reduced the data in the above way.
The candidate is detected in both K (as with NACO) and 
$L'$, where we obtain $L'=11.7 \pm 0.3$ mag by comparing
the companion candidate with the primary.\footnote{L$=6.05 \pm 0.13$ 
mag from Glass \& Penston (1974)
and Hughes et al. (1994); hence, $L' \simeq 6.05$ mag; 
the difference between $L$ and $L'$ for K7 is marginal, within the
0.3 mag error given of the companion.}
The primary has A$_{\rm V} = 0.4 \pm 0.2$ mag (Batalha et al. 2001),
hence A$_{\rm K} = 0.04 \pm 0.02$ and A$_{\rm L'} \simeq 0.02 \pm 0.01$ mag
(Rieke \& Lebofsky 1985).
Applying this extinction to the companion, we derive 
$(K-L')_{0} \simeq 1.4 \pm 0.3$ mag, consistent with L2-7
(Stephens et al. 2001, Golimowski et al. 2004).\footnote{The primary 
has a K-band excess due to its disk of $\sim 2$ mag (from a K7V 
blackdody fit to its UBVRIJHK mags).}
At $\sim 140$ pc, both the primary and the companion candidate are
by $\sim 3$ mag brighter than main-sequence K7V stars and 
Gyr old (main-sequence) early to mid L-dwarfs as far as the absolute 
magnitudes are concerned, consistent with both being young.

GQ Lup and its companion candidate are also detected in archived
images obtained with the Hubble Space Telescope (HST)
Wide Field Planetary Camera No. 2 (WFPC2)
in program SNAP 7387 (retrieved by us from the public MAST archive at STScI)
in the filters F606W and F814W. Data reduction was performed 
for the HST data as in Neuh\"auser et al. (2002). 
Since the companion candidate is detected only marginally, 
we cannot obtain reliable optical colors.

\begin{table}
\begin{tabular}{llrl}
\multicolumn{4}{c}{\bf Table 1. Observing log, FWHM, and separations} \\ \hline
Tel./Instr. & Obs. Date & FWHM & Separation \\ \hline
HST/PC      & 10 Apr 1999  & 95 mas  & $739 \pm 11$ mas \\
Subaru/CIAO & 17 Jul 2002 & 150 mas & $736.5 \pm 5.7$ mas \\ 
VLT/NACO    & 25 Jun 2004 & 68 mas  & $732.5 \pm 3.4$ mas \\
VLT/NACO    & 25 Aug 2004 & 74 mas  & $731.4 \pm 4.2$ mas \\ 
VLT/NACO    & 14 Sep 2004 & 72 mas  & $735.8 \pm 3.7$ mas \\ \hline
\end{tabular}
\end{table}

\subsection{Astrometry}

To study whether primary and companion candidate form a common
proper motion pair, we compare separations and position angles PA
at the different epochs.
We use $21.33 \pm 0.02$ mas/pixel as CIAO pixel scale
(Fukagawa et al. 2003, 2004) and 
$45.545 \pm 0.005$ mas/pixel for HST/PC (Holtzman et al. 1995).
The positions of primary and companion are determined by Gaussian
centering, after primary PSF subtraction as far as the companion
is concerned, and the errors given include the pixel scale error.
The separations measured are given in Table 1 ($1\sigma$ errors)
including the aquisition images for the spectroscopy, 
obtained with the NACO S27 camera with $27.07 \pm 0.11$ mas/pixel.
The astrometric precision is best in our own deep NACO image
(June 2004)
obtained with the S13 camera with $13.23 \pm 0.05$ mas/pixel.
Scale and orientation of S13 and S27 were obtained by us using $\sigma$ Ori AD
(Hipparcos: $12.980^{\prime \prime}$ at $84.2 ^{\circ}$)
and GJ 852 AB taking into account the orbital motion within them.
The position angle of the detector was
tilted by $0.14 \pm 0.25 ^{\circ}$ to the east.

\begin{figure}
\includegraphics[width=\columnwidth]{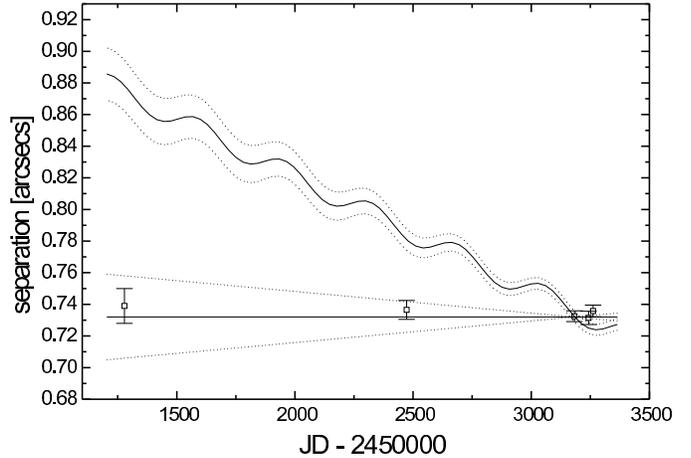}
\caption{Observed separation between primary star GQ Lup A and companion
candidate with HST/PC (left), Subaru/CIAO (middle) and VLT/NACO
(right, first the June 2004 image, then the Aug and Sept 2004 
aquisition images, all consistent within $1\sigma$) 
compared to the expectation when bound (no change in separation as
straight full line $\pm 5$ mas/yr for orbital motion as straight dotted 
lines) or when background (motion of 
GQ Lup A alone to the SW due to its proper motion $\pm$ its error with 
wobble due to expected parallactic motion of 7.1 mas).
We can reject the background hypothesis by 4.9$\sigma$ (NACO-CIAO)
plus 7.2$\sigma$ (NACO-HST).}
\end{figure}

If the faint object would be
a background object, it would have negligible
proper motion and we would see a change in separation due to
the proper and parallactic motion of GQ Lup A (Fig. 2).
Its proper motion is $\mu _{\alpha} = -27 \pm 3$ mas/yr 
and $\mu _{\delta} = -14 \pm 3$ mas/yr 
(Teixeira et al. 2000, Camargo et al. 2003)
at $140 \pm 50$ pc distance.
The background hypothesis can be rejected by 4.9 and 
7.2$\sigma$ by comparing NACO with CIAO and HST, respectively.
The PA of the companion is $275.45 \pm 0.30 ^{\circ}$ in the
NACO image in June 2004 (detector orientation calibrated);
if the candidate would be a non-moving background object, 
the PA should have been $269.89 \pm 1.03 ^{\circ}$ at the time
of the HST epoch, but we observe $275.62 \pm 0.86 ^{\circ}$ 
(detector orientation from fits file header, 
known to be stable and precise), 
i.e. $4.3 \sigma$ deviant from the background hypothesis.
Hence, the faint object is co-moving with GQ Lup A.
The probability to find by chance a fore- or background
M or L dwarf with the same proper motion near GQ Lup A
is negligible.

\subsection{Spectroscopy of the companion}

To check whether the companion is indeed cool, we
obtained a K-band spectrum with VLT/NACO using 
the {\it S54-4-SK} mode from 1.79 to $2.57~\mu$m 
with resolution 700 and nodding along the 172 mas slit.
Two spectra were obtained on 25 Aug and 14 Sept 2004 
with 20 and 40 spectra, respectively, exposed for 30s each 
and reduced in the normal way: dark subtraction, flat devision, 
shift+add, and wavelength calibration (Fig. 3).

We confirmed that the faint object was positioned well
in the center of the slit in every individual spectrum.
Wavelength dependant refraction and Strehl-ratios through the 
narrow slit can change the slope of the spectrum: 
The light-loss due to refraction,
following Schubert \& Walterscheid (2000) with the given airmasses,
parallactic and positional angle, is 
1.5 and $1.6 \%$ at the 1.9 and $2.5 \mu$m, respectively, compared to 
the flux at $2.35 \mu$m. Given the seeing and slit-width, the slit
efficiency is $76.4 \%$ at $1.67 \mu$m and $79.0 \%$ at $2.18 \mu$m,
so that the flux in the blue is reduced by $2.5 \%$.
In total, $2.0 \%$ of the flux is missing in the blue
compared to the middle of the spectrum and only $0.8 \%$ in the red.
The slit-loss is partly compensated by higher Strehl-ratio,
because the companion flux peaks in the middle of the K-band.

The flux of the companion is calibrated once with the spectrum 
of the K7eV primary star GQ Lup A observed simultaneous
and once with the telluric standard HD 159402 (B3III),
observed in the same night.
The large telluric absorption features below 2.06 $\mu$m 
and above 2.42 $\mu$m nevertheless cause a considerable
amount of noise in this part of the spectrum.
The Reid et al. (2001) K1 spectral index is 0.13 to 0.39 
giving M9-L3. The McLean et al. (2003) H$_{2}$O-D ratio 
is 0.67 to 0.89 indicating L2-7 (c.f. Cushing et al. 2005).
The NaI doublet has W$_{\lambda}=4.0 \pm 0.5$ \AA~for the primary, 
but $\le 2$ \AA~for the companion (possibly partly telluric), 
which implies a spectral type $\sim$ M9 or later 
(Cushing et al. 2005, Comer\'on et al. 2000)
or $\sim$ L2 or later (Gorlova et al. 2003);
even for strongly reddened M stars, the NaI line should still be strong 
(Greene \& Lada 1996).
The CO band head at $2.295 \mu$m is also present in M-dwarfs,
increases slightly in strength for early L, but weakens again
in later types. This band head is clearly visible with 
CO-index $\sim 0.86$, i.e. M6 to L0 (Gorlova et al. 2003).
The average of all the above estimates is L$1.5 \pm 2.5$, or M9 to L4.
This is consistent with the dereddened $K-L'$ color;
hence no evidence for additional extinction due to,
e.g., a disk around the companion.

\begin{figure}
\vspace{-1cm}
\includegraphics[width=9.8cm,height=9cm]{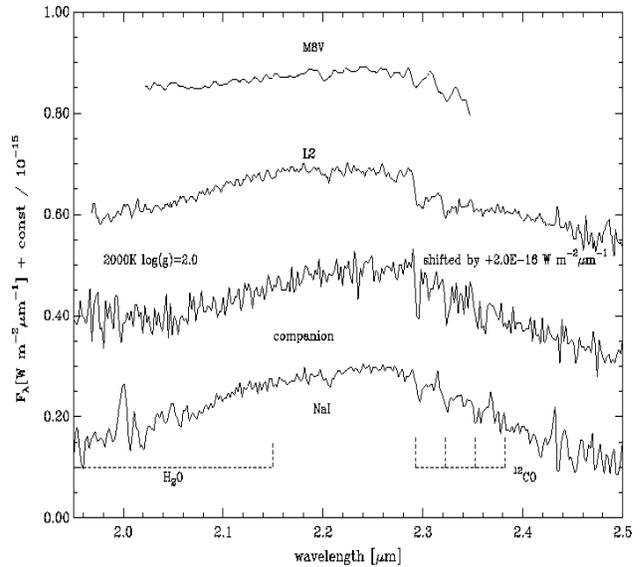}
\caption{Our flux-calibrated K-band NACO spectrum of 
the companion of GQ Lup (bottom) compared to 
the young M8 brown dwarfs in Cha I (top) from Comer\'on et al. (2000),
L2 (2MASSW J0829066+145622, 2nd from top) from Reid et al. (2001),
and a GAIA-Dusty template spectrum for 2000 K and $\log~g$=2,
which compares well with the companion.}
\end{figure}

\section{Mass determination and discussion}

Our companion, bound to a $1 \pm 1$ Myr young star, may be the
youngest and lowest-mass companion ever imaged.
Hence, it is difficult to compare it to field (old) L-dwarfs.
From the K-band CO-index being $0.862\pm0.035$
we obtain the gravity $\log~g = 2.52 \pm 0.77$,
a slight extrapolation from late-M to early-L (Gorlova et al. 2003),
hence uncertain.
Since neither spectral type nor gravity are well constrained,
the temperature T$_{\rm eff}$ is only weakly constrained to 
$\sim 1600$ to $2500$~K (or $2050 \pm 450$ K)
from Gorlova et al. (2003) and references therein
as well as Reid et al. (1999), Stephens et al. (2001), Burgasser et al. 
(2002), Nakajima et al. (2004), and Golimowski et al. (2004).

We compared our observed spectrum with the theoretical template spectra 
from the GAIA-Dusty model (Brott \& Hauschildt, in prep.), 
updated from Allard et al. 2001, with improved molecular dissociation 
constants, more dust species with opacities, spherical symmetry, 
and a mixing length parameter $1.5 \cdot $H$_{\rm p}$; we tried 
T$_{\rm eff} = 2000$ and 2900 K and $\log~g=0, 2$, and 4. 
A good fit is obtained only for 2000 K and $\log~g = 2$ (Fig. 3). 
Because the fit for $\log~g=0$ is much worse than for $\log~g=4$,
where the continuum at 2.22 to $2.3~\mu m$ and the depth of the
CO lines are not reproduced,
and because of the $\log~g$ from the CO-index (see above),
we conclude $\log~g \simeq 2$ to 3.
The observed flux can be reproduced for an object 
with $\sim 2$~R$_{\rm jup}$ radius at 140 pc.

With B.C.$_{\rm K}=3.3 \pm 0.1$ mag 
for M9-L4 and K-L'$\simeq 1.4 \pm 0.3$ (Golimowski et al. 2004),
K$_{\rm s}=13.1 \pm 0.2$ mag 
for the companion, and $140 \pm 50$ pc distance,
the luminosity is 
$\log (L/L_{\odot}) = -2.37 \pm 0.41$ 
for the companion. We can now plot it into
an H-R diagram to compare with models.
From Burrows et al. (1997) Fig. 7, 
the companion has $\sim 12$ to 32
M$_{\rm jup}$ (from L and age), 
but only $\sim 3$ to 9 
M$_{\rm jup}$ from Figs. 9 \& 10 (from T and age).
It is similar in Baraffe et al. (2002) Fig. 2: 
The companion has $\sim 3$ to 16
M$_{\rm jup}$ from T and age, 
but $\sim 11$ to 42 
M$_{\rm jup}$ from L and age;
$\sim 12$ to 42 
M$_{\rm jup}$ from M$_{\rm K}$ and age
(see also perso.ens-lyon.fr/isabelle.baraffe).
Burrows et al. and Baraffe et al. 
start with an assumed internal structure without collapse,
so that their models are uncertain up to a few Myrs (Baraffe et al. 2002).

\begin{figure}
\includegraphics[width=\columnwidth,height=7.5cm]{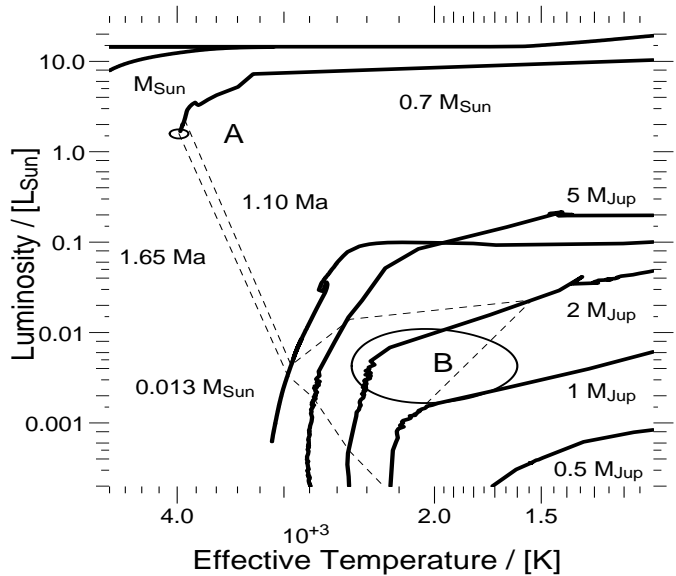}
\caption{H-R diagram with tracks from Wuchterl \& Tscharnuter
(2003) plus more tracks calculated by us (from top to bottom) 
for 1 and 0.7~M$_{\odot}$ as well as 13, 5, 2, 1, and 0.5~M$_{\rm jup}$.
The isochrones (dashed) for 1.10 and 1.65 Myr connect the end of the 
0.7~M$_{\odot}$ track (star A) with the planetary tracks.
Components A and B (1$\sigma$ errors) 
are co-eval at $\sim 1.1$ Ma.}
\end{figure}

Wuchterl \& Tscharnuter (2003) include the initial collapse (Fig. 4):
Tracks for masses of 1 to $0.013$~M$_{\odot}$ are radiation fluid-dynamical 
calculations of the collapse of initially marginally unstable Bonnor-Ebert-spheres;
planetary tracks are models in the framework of the nucleated instability 
hypothesis (Wuchterl et al. 2000; Wuchterl, in preparation). 
All ages are counted from first photosphere formation.
The age-offset visible on the 1.10 Myr isochrone, at the transition from the
0.013~M$_{\odot}$ brown dwarf collapse to the 5~M$_{\rm jup}$ planetary track 
is due to the time-offset caused by the planetary core 
formation
($\le 10^{5}$ yrs).
The companion has a mass 
of $\sim 1$ to 2~M$_{\rm jup}$ (1$\sigma$) 
and is co-eval with the star ($\sim 1.1$~Myr).
From this model (Fig. 4), we expect $\log~g \simeq 2.4$
and $\sim 1.8$~R$_{\rm jup}$ radius, consistent with 
our spectrum and the best fitting GAIA-Dusty model spectrum
(Fig. 3).

The most critical point in the mass determination of the companion
(candidates) of GQ Lup and 2M1207 are the models, 
which may be off by an unknown factor for low ages (few Myrs);
they need to be calibrated, before the mass of such companions
can be determined confidently.

\acknowledgements{We thank N. Reid for his IR spectra of L-dwarfs,
an anonymous referee for very good suggestions,
and the ESO USG and Paranal teams for perfect support.
RN would like to thank DLN for long-term support.}

{}

\end{document}